\documentclass{aa}  

\usepackage{graphicx}
\usepackage{txfonts}
\begin{document}

   \title{Detection of the secondary eclipse of Qatar-1b in the Ks band \thanks{Based on observations collected at the Calar Alto Observatory, Almer\'ia, Spain.}}

   \author{Patricia Cruz \inst{1,2}, David Barrado \inst{2},
   		Jorge Lillo-Box \inst{3,2}, Marcos Diaz \inst{1}, Jayne Birkby \inst{4,5},
			Mercedes L\'opez-Morales \inst{4}, Jonathan J. Fortney \inst{6}
          }

   \institute{Instituto de Astronomia, Geof\'isica e Ci\^encias Atmosf\'ericas, Universidade de S\~ao Paulo (IAG/USP), S\~ao Paulo, Brazil\\
              \email{patricia.cruz@usp.br}
         \and
             Depto. de Astrof\'isica, Centro de Astrobiolog\'ia (INTA-CSIC), ESAC campus, Camino Bajo del Castillo s/n, E-28692, Villanueva de la Ca\~nada, Spain
         \and
             European Southern Observatory (ESO), Alonso de Cordova 3107, Vitacura, Casilla 19001, Santiago de Chile, Chile
         \and
             Harvard-Smithsonian Center for Astrophysics, 60 Garden Street, Cambridge, MA 02138, USA
         \and
             NASA Sagan Fellow
         \and
             Department of Astronomy and Astrophysics, University of California, 1156 High Street, Santa Cruz, CA 95064, USA
             }

   \date{Received XXXXX; accepted XXXXX}

 
  \abstract
   {}
   {Qatar-1b is a close-orbiting hot Jupiter ($R_p\simeq 1.18$ $R_J$, $M_p\simeq 1.33$ $M_J$) around a metal-rich K-dwarf, with orbital separation and period of 0.023 AU and 1.42 days. We have observed the secondary eclipse of this exoplanet in the Ks band with the objective of deriving a brightness temperature for the planet and providing further constraints to the orbital configuration of the system.}
   {We obtained near-infrared photometric data from the ground by using the OMEGA2000 instrument at the 3.5 m telescope at Calar Alto (Spain) in staring mode, with the telescope defocused. We have used principal component analysis (PCA) to identify correlated systematic trends in the data. A Markov chain Monte Carlo analysis was performed to model the correlated systematics and fit for the secondary eclipse of Qatar-1b using a previously developed occultation model.
   We adopted the prayer bead method to assess the effect of red noise on the derived parameters.
   }
   {We measured a secondary eclipse depth of $0.196\%^{+0.071\%}_{-0.051\%}$, which indicates a brightness temperature in the Ks band for the planet of $1885^{+212}_{-168}$ K. We also measured a small deviation in the central phase of the secondary eclipse of $-0.0079^{+0.0162}_{-0.0043}$, which leads to a value for $e\cos{\omega}$ of $-0.0123^{+0.0252}_{-0.0067}$. However, this last result needs to be confirmed with more data.
   \thanks{The light-curve data shown in Fig.3 are only available in electronic form at the CDS via anonymous ftp to cdsarc.u-strasbg.fr (130.79.128.5) or via http://cdsweb.u-strasbg.fr/cgi-bin/qcat?J/A+A/.}
   }
{}


\authorrunning{P. Cruz et al.}
\titlerunning{Detection of the secondary eclipse of Qatar-1b in the Ks band}

   \maketitle
%

\section{Introduction}\label{intro}

Qatar-1b is a close-orbiting hot Jupiter, with an initially measured planetary radius of $R_p= 1.164\pm0.045$ $R_J$ and a mass of $M_p= 1.090^{+0.084}_{-0.081}$ $M_J$ (Alsubai et al. 2011). It orbits its central star, a metal-rich K-dwarf, with a period of approximately $1.42$ days and at an orbital separation of $0.02343^{+0.00026}_{-0.00025}$ AU. This exoplanet was first assumed to be in a circular orbit (fixed $e=0.0$; Alsubai et al. 2011). 

Covino et al. (2013) published five new transit epochs, which combined with previous data from \cite{Alsubai11} resulted in new ephemerides and improved parameters for the Qatar-1b system ($R_p= 1.18\pm0.09$ $R_J$; $M_p= 1.33\pm0.05$ $M_J$). These authors also presented a revised orbital solution for this object, based on 11 radial velocity (RV) measurements, finding an eccentricity of $e=0.020^{+0.011}_{-0.010}$. They also observed the Rossiter-McLaughlin effect in the RV curve, determining the sky-projected obliquity and concluding that the system is well aligned.

In addition, \cite{vonEssen13} reported indications of long-term transit-timing variations (TTVs) based on 26 transits. To explain the observed $\sim$$190$-day TTV period, these authors presented different cases that could be responsible for such variations. One of the possibilities is a weak perturber in resonance with Qatar-1b, since two planetary bodies in resonance would experience long-term variations in their orbital parameters. In contrast, very recent works by \cite{Mislis15} and \cite{Maciejewski15} did not find conclusive evidence of TTVs. Nevertheless, \cite{Mislis15} indicated that more precise data are needed for further conclusions.

In this paper we present the second result of the Calar Alto Secondary Eclipse study (the CASE study; see Cruz et al. 2015): a detection of the secondary eclipse of Qatar-1b in Ks band, with the goal of measuring its Ks-band thermal emission for the first time. We also aim to constrain the planet's orbital configuration by determining offsets in the mid-eclipse timing that would indicate a non-circular orbit and possible perturbations. 
This paper is structured as follows: in Sect. 2 we present the data acquisition and the reduction procedures applied to them; in Sect. 3 we describe our analysis, including the correction for systematic effects, the modeling of the eclipse, and the fitting results; in Sect. 4 we discuss the thermal emission and the orbital configuration of Qatar-1b; and in Sect. 5 we present our conclusions.


\section{Observations and data reduction}\label{obsred}

Qatar-1 ($K$=$ 10.409$ mag) was observed in service mode on the night of 2011 August $30$, under photometric conditions. This night was selected by estimating the occurrance of a secondary eclipse, assuming a circular orbit\footnote{The secondary eclipse timing, assuming a circular orbit, was predicted with the help of the Exoplanet Transit Database, ETD, maintained by the Variable Star Section of Czech Astronomical Society - for more details, see http://var2.astro.cz/ETD/index.php.}. We used the Ks-band filter (centered on 2.14 $\mu $m) of the OMEGA2000 instrument equipped with a 2k x 2k HAWAII-2 detector, on the 3.5 m telescope at the Calar Alto Observatory (CAHA) in southern Spain. The field of view is of 15.4 x 15.4 arcmin and the plate scale of 0.45 arcsec pix$^{-1}$. To reduce intrapixel variations and minimize the effect of flat-field errors, the telescope was arbitrarily defocused, resulting in a ring-shaped point spread function (PSF) with a mean radius of $\sim $$4.45$ arcsec (about $10$ pixels). The relative photon noise in a single raw defocused image is of $\sim$$4.6\times 10^{-3}$.

We acquired the data in staring mode, observing the target continuously without dithering. This technique has been used for similar purposes by several authors, for example, Croll et al. (2010a, 2010b, 2011), de Mooij et al. (2011), and Cruz et al. (2015). We gathered a series of data where every file has 15 images of 4s exposure each ($190\times15\times4$${\rm s}+$ overheads), 2850 individual measurements in total, with an average cadence of $\sim$$11.3$ images per minute. A manual guiding correction was performed, keeping the stars on the same position on the detector as best possible. These staring mode observations were collected during approximately 4.2 hours. Before and after this sequence, in-focus images composed by five dither-point images each were also obtained for further sky subtraction. 
The data reduction was performed using IRAF\footnote{IRAF is distributed by the National Optical Astronomy Observatory, operated by the Association of Universities for Research in Astronomy, Inc., under cooperative agreement with the National Science Foundation.} for the bad pixel removal, flat-fielding, and sky subtraction, which are described in detail in \cite{Cruz15}.

We performed aperture photometry in each image using the {\it aper} procedure from the IDL Astronomy User's Library\footnote{IDL stands for Interactive Data Language - for further information, see http://www.ittvis.com/ProductServices/IDL.aspx; {\it aper.pro} is distributed by NASA - see http://idlastro.gsfc.nasa.gov/ for more details.}. We selected a circular aperture with a radius of 12.5 pixels and inner and outer annuli of 19.5 and 36.5 pixels, respectively, to measure the residual sky background. For the photometry, we tested different apertures from 3 to 25 pixels, in steps of 0.5 pix. We used the aperture that presented the highest signal-to-noise ratio for the target. Sky annuli were also varied by several pixels, but had negligible effect on the final photometry.

Table \ref{refstars2MASS} shows the stars used as reference for the differential photometry. 
These stars were selected because they did not present any strong variations or other odd behavior in their light curves.

\begin{table}
\caption{Reference stars used for the relative photometry.}\label{refstars2MASS}
\centering
\small
\begin{tabular}{lccc}
\hline\hline
Star No. & Identifier (2MASS) & K mag \\
\hline
1 & J20142800+6506270 & 9.231 \\
2 & J20131440+6507000 & 10.560 \\
3 & J20142873+6507445 & 9.584 \\
4 & J20123648+6508138 & 9.254 \\
5 & J20141068+6511408 & 10.674 \\
6 & J20133795+6512408 & 10.643 \\
7 & J20140978+6512548 & 8.995 \\
8 & J20125213+6514365 & 9.347 \\
9 & J20135707+6515082 & 8.879 \\
\hline
\end{tabular}
\end{table}

\begin{figure}
   \centering
   \includegraphics[width=1.0 \columnwidth,angle=0]{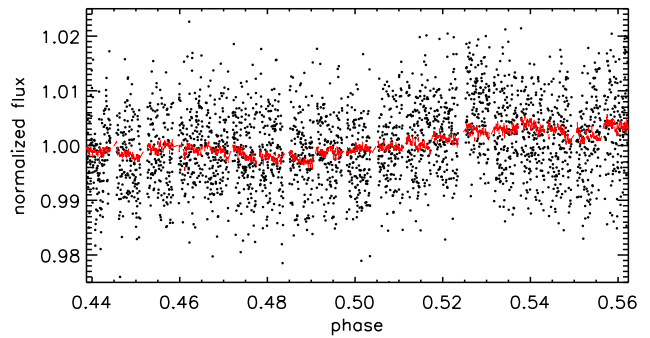}
      \caption{Uncorrected differential light curve of Qatar-1 in the Ks band shown in phase. The thin line represents the best model obtained from the joint-MCMC.}
         \label{LCuncorrected}
   \end{figure}

The relative flux of Qatar-1 was calculated by its measured flux, $F_{\rm tar}(t)$, over the sum of the flux of all reference star considered, and then normalized by its median value, as follows:
   \begin{equation}\label{eqflux}
		F(t) = \dfrac {F_{\rm tar}(t)} {\sum_{i=1}^{9}F_{\rm ref,i}(t)},
   \end{equation}
   \begin{equation}\label{eqflux2}
		f(t) = \dfrac {F(t)} {\tilde{F}} ,
   \end{equation}
where $F_{\rm ref,i}(t)$ is the measured flux of reference star $i$, at a given time $t$, $\tilde{F}$ is the median value of the target relative flux estimated from the expected out-of-eclipse section of the light curve, and $f(t)$ is the normalized flux of the target. 
The differential light curve of Qatar-1 is shown in Fig. \ref{LCuncorrected}.


\section{Data analysis and results}\label{analys}

The analysis of the data presented here followed the steps presented in the previous paper of this series (Cruz et al. 2015). We maintained the same procedure to obtain a homogeneous analysis, for coherent results.

\subsection{\bf Correction of systematic effects}\label{syseff}

To clean the light curve from systematic effects, we searched for correlations between the observed flux and several other parameters. We used the principal component analysis (PCA) technique, a powerful statistical tool for dimensional evaluation in data sets, to identify the systematics that have a direct effect on the data (for more details, see Cruz et al. 2015). In a $R^{n}$ array, the PCA finds the linear combination (vector) of ${n}$ axes that best reproduces the data distribution in question (for more, see Morrison 1976). 
This method was primarily used to recognize which variables are responsible for most of the variance in the data set. The purpose here was to identify the parameters that strongly affect the data. It is worth emphasizing, at this point, that the data were not yet corrected for systematics. The parameters with a high eigenvalue were selected for a posterior elimination of the dominant patterns without compromising the eclipse signal.

We calculated the PCAs for a few reference stars listed in Table \ref{refstars2MASS}. Those stars with similar 2MASS colors were selected to minimize differential refraction and other chromatic effects. They were treated separately, with their normalized fluxes obtained from Eqs.\ref{eqflux} and \ref{eqflux2}, where ${F_{\rm tar}}$ is the measured flux of the reference star. We analyzed different reference light curves, where we excluded the star in question from the sum in Eq.\ref{eqflux} when dealing with a particular reference star. 
These analyses revealed significant correlations of the normalized flux with a star's xy-position of the centroid at the detector ($x_{c}$, $y_{c}$), ``defocused seeing'' ($s$), and airmass ($\sec z$). 
{Since we used a fixed aperture (see Sect. \ref{obsred}) for the whole data set, a correction was estimated by defining a new radius around the target centered from a boundary of four sigma above the residual background for each image individually. From this new radius, we obtained the ``defocused seeing'' by multiplying it by the plate scale.}

The behavior of these parameters as function of the orbital phase is presented in Appendix \ref{appen}. The light curves of the reference stars were considered only to identify the systematics and were not used to analyze the light curve of Qatar-1. We also calculated the PCAs for Qatar-1 and found correlations with the same parameters. For this, we considered only the expected out-of-eclipse (ooe) part of the light curve, assuming a circular orbit, where the stellar flux is supposed to be constant.

Thereafter, we performed a multiple linear regression in IDL ({\it regress}), fitting for these systematics simultaneously, generating a polynomial of the form:
   \begin{equation}\label{eqsys}
 		f_{\rm ooe} = c_{0} + c_{1}x_{\rm c,ooe} + c_{2}y_{\rm c,ooe} + c_{3}s_{\rm ooe} + c_{4}\sec z_{\rm ooe} .
   \end{equation}
Here, $f_{\rm ooe}$ is the out-of-eclipse flux of the target and $c_{k}$ are constants of the fit. This linear regression was performed only in the expected out-of-eclipse part of the light curve to avoid removing the eclipse signal along with systematics. This correction function was determined by using the light curve of the target alone, and was corrected point by point considering the xy-position of the target to assess the behavior of the residuals. 
From the out-of-eclipse portion of the decorrelated unbinned light curve, we estimated a root-mean-square, RMS, of $\sim$$7.1\times 10^{-3}$.

  \begin{figure}
   \centering
   \includegraphics[width=1.0 \columnwidth,angle=0]{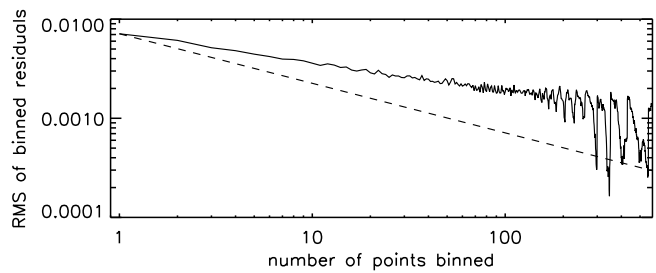}
      \caption{RMS of the residuals of the out-of-eclipse portion of the light curve for different bin sizes. The dashed line shows the limit expectation for normally distributed noise.}
         \label{rms}
   \end{figure}

We show in Fig. \ref{rms} the behavior of the residuals. The expected Gaussian noise, defined as one over the square-root of the bin size, is also presented for comparison. Considering only the expected out-of-eclipse data, we have explored the noise level (RMS) behavior by binning the curve with different bin sizes, from 1 to 580 points per bin from a total of 1761 out-of-eclipse individual measurements. In the latest case, the binned light curve would have only three points. 
The RMS of the residuals is higher than the white noise, which we attribute to the presence of significant correlated noise in the photometric data.

We analyzed the power spectrum of the data in question to estimate the red noise and search for higher frequency periodic variations in the light curve. We generated a periodogram of the detrended data with and without the eclipse (by removing the eclipse model obtained from the analysis presented in Sect. \ref{MAmodeling}), and significant red noise was detected on timescales greater than $\sim$6 minutes for both cases. The average power at eclipse frequencies is $\sim$9 times the white noise.

\subsection{\bf Model for the secondary eclipse}\label{MAmodeling}

\begin{table}
\caption{Qatar-1b parameters of the system from \cite{Covino13} used in the analysis.}\label{SysParam}
\centering
\small
\begin{tabular}{lc}
\hline\hline
Parameter & Value \\
\hline
Planet-star radii ratio $R_{p}/R_{*}$	& 0.1513$\pm $0.0008  \\
Transit epoch $T_{0}$ (days)	& 2455518.41094$\pm $0.00016  \\
Transit duration $T_{1-4}$ (days)	& 0.0678$\pm $0.0010 \\
Orbital period $P$ (days)	& 1.42002504$\pm $0.00000071  \\
Orbital inclination $i$ (deg)	& 83.82$\pm $0.25  \\
Semimajor axis $a$ (AU)	  &   0.02343$\pm $0.0012  \\
Stellar radius $R_{*}$ ($R_{\odot}$)	& 0.80$\pm $0.05  \\
Planet radius $R_{p}$ ($R_{Jup}$)	& 1.18$\pm $0.09  \\
Stellar $T_{\rm eff}$ (K)	& 4910$\pm $100  \\
\hline
\end{tabular}
\end{table}

We performed a Markov chain Monte Carlo (MCMC) analysis to fit for the Qatar-1b secondary eclipse, where we adopted the transit model from \cite{MandelAgol02}, assuming no limb-darkening. Table \ref{SysParam} shows the orbital parameters used in the occultation model and in the analysis, published by Covino et al. (2013).

An initial modeling of the detrended unbinned light curve (with all 2850 individual measurements) was performed to define the initial conditions for the MCMC analysis, where we fit for three parameters: the eclipse depth ($\Delta F$), a phase deviation ($\Delta \phi$) of the eclipse to search for variations in time, which would imply a non-zero eccentricity, and the baseline level ($F_{bl}$). This resulted in an eclipse depth, $\Delta F$, of $0.181\%$, a light-curve baseline level, $F_{bl}$, at $1.0003$, and a measured deviation in phase from mid-eclipse timing, $\Delta \phi$, of $-0.0069$.

We then performed the MCMC analysis of the uncorrected light curve, unifying the decorrelation function and the occultation model in a joint fit (hereafter joint-MCMC). This way, we were able to integrate the effect of the systematics into the uncertainty estimates, as in \cite{Cruz15}. We simultaneously fit for the secondary eclipse and the systematic effects, generating a model of the form
\begin{equation}
model=m_{\rm occ} \times f_{\rm sys} .
\end{equation}
The occultation model is represented here by $m_{\rm occ}$. The polynomial $f_{\rm sys}$ is given by Eq. \ref{eqsys}, but it is calculated for the whole light curve, including the data during the eclipse. We considered the values defined previously for the eclipse model as initial conditions, except for the baseline level, which was kept fixed at $F_{bl}=1.0$. The coefficients obtained by the linear regression (Sect. \ref{syseff}) were also used as initial conditions.

  \begin{figure}
   \centering
      \vspace*{2mm}
   \includegraphics[width=1.04 \columnwidth,angle=0]{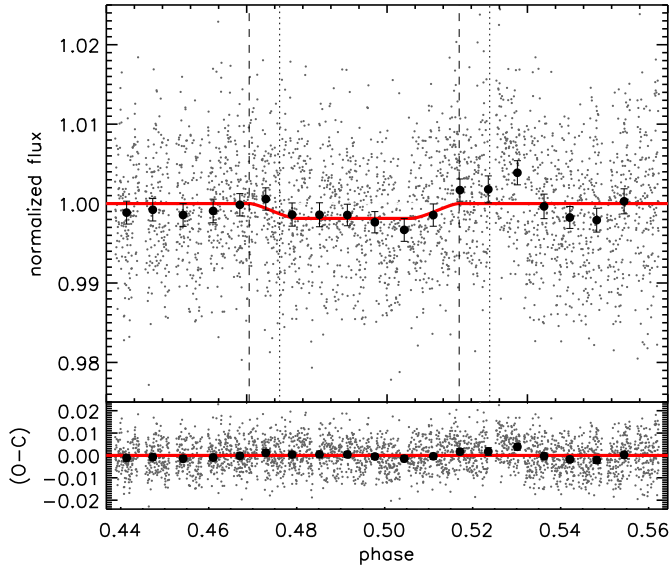}
      \vspace*{5mm}
      \caption{Secondary eclipse of Qatar-1b in the Ks band, showing the best-fitting model with $\Delta F = 0.186\%$, and $\Delta \phi = -0.0071$. The detrended light curve is presented, where the small dots show the individual measurements. The solid line represents the best occultation model obtained from the joint-MCMC analysis, after removing the contribution of the systematics. The dotted vertical lines show the ingress and egress positions expected for circular orbit, and the dashed lines present ingress and egress found in the analysis.  The filled circles show the light curve binned every 143 points ($\sim$$9.5$ minutes), for better visualization.}
         \label{MAmodelFit2}
   \end{figure}

We ran four chains of $1.1\times 10^{6}$ simulations each. The first $1\times 10^{5}$ simulations from each chain were neglected, so that the analysis is not directly influenced by the initial conditions. In total, we generated $4\times 10^{6}$ models, considering the simulations from all four chains. From this analysis, we obtained an eclipse depth of $0.186\%^{+0.022\%}_{-0.024\%}$ in flux, and a deviation in phase of $-0.0071^{+0.0006}_{-0.0010}$, given by the median in the distributions from the joint-MCMC analysis. The uncertainties were determined by the 16-84$\%$ interval of the joint-MCMC parameter distribution. 
The convergences of the MCMC chains are at $10^{-4}$ for the phase shift and at $10^{-5}$ for the depth of the eclipse.

Figure \ref{MAmodelFit2} presents the best occultation model from the joint-MCMC for the unbinned light curve, after removing the contribution of the systematics. The dotted vertical lines show the ingress and egress positions expected for the secondary eclipse considering a circular orbit, and the dashed lines present the deviation in phase found in the analysis. 

Since the red noise present in the data is not negligible (see Sect. \ref{syseff}), we used the prayer bead method (Moutou et al. 2004, Gillon et al. 2007) to assess the effect of red noise on the derived parameters. We revised the uncertainties from the distribution of the values obtained from MCMC simulations with the prayer bead method (hereafter PB-MCMC). Here, we first shifted sequentially the residuals of the joint-MCMC fit and then added the best-fit occultation model. In this way, we generated as many light curves as residual permutations that went through the same MCMC procedure.

From the distribution of values obtained from the PB-MCMC, we derived an eclipse depth of $0.196\%^{+0.071\%}_{-0.051\%}$ and a phase shift of $-0.0079^{+0.0162}_{-0.0043}$ ($-16.15^{+33.12}_{-8.79}$ minutes). Again, the uncertainties were determined by the 16-84$\%$ interval of the PB-MCMC parameter distribution. 
We note that we have a limited sample of red noise, and its timescale is comparable with the eclipse duration, which could compromise the shape of the distributions obtained from the PB-MCMC analysis and may yield an overestimation of the derived uncertainties.


\section{Discussion}\label{discus}

  \begin{figure}
   \centering
   \includegraphics[width=1.0 \columnwidth,angle=0]{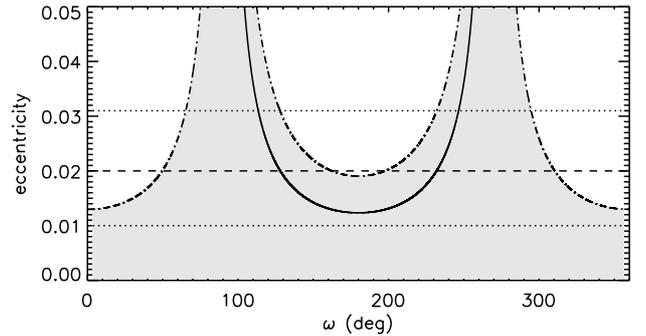}
      \caption{Distinct orbital configurations for $e\cos{\omega}=-0.0123^{+0.0252}_{-0.0067}$, with the value $e\cos{\omega}=-0.0123$ presented as a solid line. The lower and upper limits are shown as dash-dotted lines. The whole interval of $-0.0190 \leq e\cos{\omega} \leq 0.0129$ is represented by the gray shaded area. The horizontal dotted lines show the eccentricity range found by \cite{Covino13}, with $e=0.02$ presented as a dashed horizontal line.}
         \label{ecc}
   \end{figure}

\subsection{Derived orbital configuration}\label{sececc}

Based on the observation of a phase offset, it was possible to investigate possible intervals for the eccentricity, $e$, and the argument of periastron, $\omega $, by following the relationship (Wallenquist 1950, L\'opez-Morales et al. 2010)
   \begin{equation}
		e\cos{\omega} =  \pi \cdot \dfrac {\delta \phi - 0.5} {1 + \csc^{2}(i)},
   \end{equation}
where $\delta \phi = \phi_{ecl}-\phi_{tra}$, and $i$ is the orbital inclination of the system. 
The phase of the secondary eclipse is defined as $\phi_{ecl}=0.5+\Delta \phi$.

We derived a phase offset for the Qatar-1b system of $\Delta \phi = -0.0079^{+0.0162}_{-0.0043}$, which combined with the orbital inclination of Qatar-1b (shown in Table \ref{SysParam}) results in a value for $e\cos{\omega}$ of $-0.0123^{+0.0252}_{-0.0067}$, from the equation above. This value of $e\cos{\omega}$ can lead to several combinations of $e$ and $\omega $ that are consistent with a near-to-circular orbit configuration.

Figure \ref{ecc} presents possible $e$ and $\omega $ combinations suitable for the estimated $e\cos{\omega}$, where the value $e=-0.0123$ is shown as a solid line and the lower and upper limits are presented as dash-dotted lines. All possible combinations are represented by the gray shaded area. With high-precision RV measurements, \cite{Covino13} determined a low eccentricity of $e=0.020^{+0.011}_{-0.010}$. Their result is shown as dashed and dotted horizontal lines in Fig. \ref{ecc}, as the range of possible eccentricities ($ 0.010 \leq e \leq 0.031 $). Based on our derived $e\cos{\omega}$ and its uncertainties, it is not possible to constrain $\omega$ for eccentricities lower than $0.02$.

It is worth noting that the derived $e\cos{\omega}$ in this work need to be further investigated with additional data.

\subsection{Thermal emission of Qatar-1b}\label{thermal}

From the secondary eclipse depth we can also derive the brightness temperature of Qatar-1b in the Ks band, $T_{\rm Ks}$. As presented in Sect. \ref{MAmodeling}, we measured an eclipse depth (i.e., flux ratio at a given wavelength) of $\Delta F=0.196\%^{+0.071\%}_{-0.051\%}$. 
Considering the parameters in Table \ref{SysParam} and assuming that Qatar-1b and its central star emit as blackbodies, we can estimate a Ks-band brightness temperature of $T_{\rm Ks}\simeq 1885^{+212}_{-168}$ K.

Assuming an inefficient energy circulating atmosphere ($f=2/3$) and a zero Bond albedo ($A_{B}=0$), we would expect a maximum equilibrium temperature\footnote{$T_{eq} = T_{s}\cdot({R_{s}}/{a})^{1/2}\cdot[f\cdot(1-A_{B})]^{1/4}$. $T_{s}$ and $R_{s}$ are the stellar effective temperature and radius, $a$ is the orbital semimajor axis, $f$ is the energy reradiation factor, and $A_{B}$ is the Bond albedo of the planet.} for this exoplanet of $T_{eq}\simeq 1768$ K. In this case, the derived Ks-band brightness temperature ($T_{\rm Ks}$) of Qatar-1b agrees with the maximum estimated $T_{eq}$ within the error estimates.

Some studies have suggested that an opacity source, for instance, TiO and VO, in a high layer of a planet atmosphere can lead to a hot stratosphere and hence to a temperature inversion layer. According to \cite{Fortney06}, atmospheric models with a temperature inversion would show a weak emission in the near-infrared (JHK bands), and stronger emissions would be present in models without such an inversion layer at the same wavelength bands. The measured planet-to-star flux ratio, given by the eclipse depth, was then compared with several atmospheric spectral models by Fortney et al. (2006, 2008). These models were generated considering different reradiation factors ($f$), and computed without TiO/VO, since such models tend to have higher K-band fluxes.

  \begin{figure}
   \centering
   \includegraphics[width=0.98 \columnwidth,angle=0]{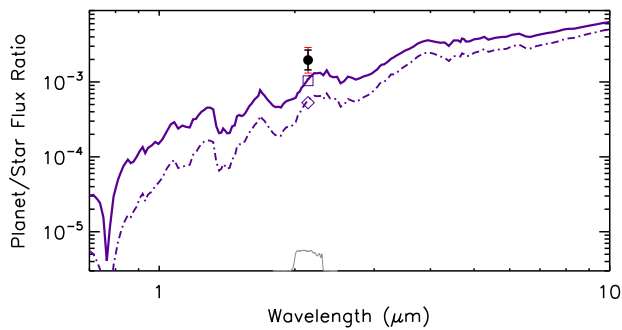}
      \caption{Model spectrum of thermal emission of Qatar-1b with a reradiation factor of $f=2/3$, without TiO/VO (solid line). The filled circle shows the measured planet-to-star flux ratio in the Ks band ($\lambda_c$=2.14 $\mu $m). The error bar shown in red represents the expected flux ratio for a temperature range of $T_{\rm Ks}\pm200$ K. The dot-dashed line illustrates a similar model with $f=1/2$. The Ks-band transmission curve (gray line) is shown at the bottom of the panel at arbitrary scale.}
         \label{spectra}
   \end{figure}

Figure \ref{spectra} shows as an example the atmospheric spectral model that better represents the observed thermal emission of Qatar-1b. This model, shown as a solid line, was generated considering an instant reradiation over the dayside ($f=2/3$), and it was calculated without TiO/VO. Although it cannot reproduce the measured planet-to-star flux ratio (filled circle), the high observed $T_{\rm Ks}$ suggests an extremely inefficient redistribution onto the night side, or missing physics in the atmosphere model, since the planet is hotter than the hottest ($f=2/3$) model.

The observed emission may suggest that Qatar-1b does not have a temperature inversion, if we assume the absence of TiO and VO in its atmosphere. Nevertheless, more observations are needed, at different wavelengths, to allow a complete understanding of this exoplanet's atmosphere and to confirm or exclude the absence of an inversion layer.

Stellar activity may also explain the high observed brightness temperature of Qatar-1b. \cite{Khodachenko07} reported that coronal mass ejections can interact with upper planetary atmospheres and may affect their temperature structure. In addition, constant UV flares and X-ray radiations may heat the exoplanet atmosphere and enhance atmospheric escape (Khodachenko et al. 2007, Lammer et al. 2007). Considering the proximity of Qatar-1b to its host star ($0.02343$ AU), such interactions should not be neglected.

\cite{Mislis15} searched for signs of activity in the available photometric database. Since starspots are common on K dwarfs, and may be responsible for a brightness modulation at the stellar rotational period, the authors analyzed the survey photometry in Qatar-1b discovery paper to search for signs of spot activity. Their results showed a sinusoidal variation in the light curve that was interpreted to be due to the presence of spots. However, there was no evidence of starspots through occultation during the observed transits.

Hence, further analysis of the stellar activity is needed. The effect of this activity also needs to be investigated to characterize the atmosphere of Qatar-1b.


\section{Conclusions}\label{concl}

We have observed the secondary eclipse of Qatar-1b in the Ks band. The observations were made in staring mode at the 3.5 m telescope at Calar Alto (Spain), equipped with the OMEGA2000 instrument.

The obtained light curve was treated with the same procedure as described in \cite{Cruz15}. 
We used principal component analysis to identify correlated systematic trends in the data. The following step was to perform a Markov chain Monte Carlo analysis to model the correlated systematics and fit for Qatar-1b secondary eclipse using the transit model of \cite{MandelAgol02}. From this analysis, we obtained an eclipse depth ($\Delta F$) of $0.186\%^{+0.022\%}_{-0.024\%}$ in flux and a deviation in phase from mid-eclipse ($\Delta \phi $) of $-0.0071^{+0.0006}_{-0.0010}$. We adopted the prayer bead method to assess the effect of red noise on the derived parameters, which resulted in an eclipse depth of $0.196\%^{+0.071\%}_{-0.051\%}$ and a deviation in phase of $-0.0079^{+0.0162}_{-0.0043}$.

The observed phase shift leads to a value for $e\cos{\omega}$ of $-0.0123^{+0.0252}_{-0.0067}$, which needs to be confirmed with more data. The planet-to-star flux ratio observed resulted in a brightness temperature in the Ks band of $T_{\rm Ks}\simeq 1885^{+212}_{-168}$ K, which agrees with the maximum equilibrium temperature of $T_{eq}\simeq 1768$ K, within the error estimates. Further analysis on the stellar activity and a study of the effect of this activity on Qatar-1b need to be performed to characterize the atmosphere of this close orbiting exoplanet.
%


\begin{acknowledgements}
This research has been funded by Spanish grants AYA2012-38897-C02-01, and PRICIT-S2009/ESP-1496. PC, DB, and JB have received support from the RoPACS network during this research, a Marie Curie Initial Training Network funded by the European Commissions Seventh Framework Programme. J.L-B acknowledges financial support from the Marie Curie Actions of the European Commission (FP7-COFUND) and the Spanish grant AYA2012-38897-C02-01.

This work was performed in part under contract with the California Institute of Technology (Caltech)/Jet Propulsion Laboratory (JPL) funded by NASA through the Sagan Fellowship Program executed by the NASA Exoplanet Science Institute.

This article is based on data collected under Service Time program at the Calar Alto Observatory, the German-Spanish Astronomical Center, Calar Alto, jointly operated by the Max-Planck-Institut f\"ur Astronomie Heidelberg and the Instituto de Astrof\'isica de Andaluc\'ia (CSIC). We are very grateful to the CAHA staff for the superb quality of the observations. 

This work has made use of the ALADIN interactive sky atlas and the SIMBAD database, operated at CDS, Strasbourg, France, and of NASA's Astrophysics Data System Bibliographic Services.

\end{acknowledgements}


\begin{appendix} 
\section{Behavior of selected baseline parameters}\label{appen}

\begin{figure}[!h]
   \centering
      \vspace*{2mm}
   \includegraphics[width=1.04 \columnwidth,angle=0]{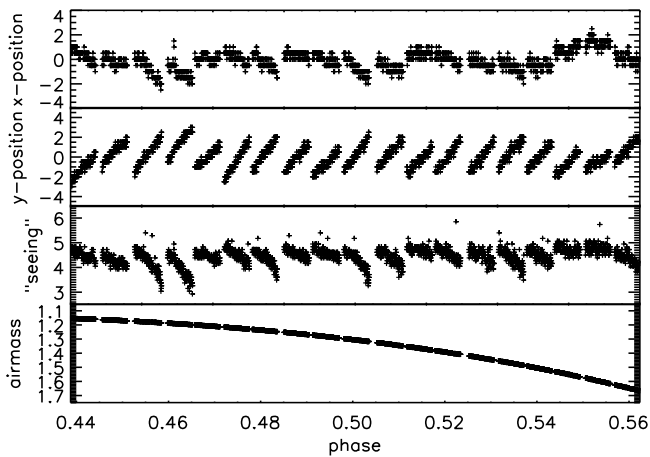}
      \vspace*{5mm}
      \caption{Behavior of the xy-position of the centroid at the detector (in pixels), the ``defocused seeing'' (in arcsec), and the airmass as a function of phase.}
         \label{baselineparam}
   \end{figure}

Figure \ref{baselineparam} shows the xy-position of the centroid at the detector, the ``defocused seeing'' (defined previously in Sect. \ref{syseff}), and the airmass as a function of phase to illustrate the behavior of the baseline selected parameters over time. Here, the first two plots show the variation in position of the star, which is in fact the centroid position minus the median of the position of the star overnight, in pixels. The seeing is presented in arcsec.

\end{appendix}

\end{document}